\documentclass[aps,floatfix,eqsecnum,showpacs,preprint]{revtex4}

\usepackage{epsfig}

\usepackage{amsmath}

\def\e{\kern+.6ex\lower.42ex\hbox{$\scriptstyle \iota$}\kern-1.20ex e}

\begin{document}

\title{The six-nucleon Yakubovsky equations for $^6He$
}

\author{W.~Gl\"ockle$^1$}

\author{H.~Wita{\l}a$^2$}

\affiliation{$^1$Institut f\"ur theoretische Physik II,
Ruhr-Universit\"at Bochum, D-44780 Bochum, Germany}

\affiliation{$^2$M. Smoluchowski Institute of Physics, Jagiellonian
University, PL-30059 Krak\'ow, Poland}

\date{\today}

\begin{abstract}
The six-nucleon problem for the bound state is formulated in
 the Yakubovsky scheme. Hints for a numerical implementation are provided.
\end{abstract}


\maketitle \setcounter{page}{1}

\section{Introduction}
\label{sec_one}

There is a rich literature on $ ^6 He $  based on an effective 
$ \alpha-n-n $ 3-body problem \cite{rep}. Besides pair interactions also 
ad hoc 3-body forces are used.
 The Pauli principle is approximately incorporated by projecting out 
"Pauli forbidden states" for the neutrons inside the $\alpha$-particle 
wave function. While that approach catches presumably the 
halo structure of the two loosely bound neutrons clearly because of
its strongly restricted ansatz
 it is not suited  to  probe modern nucleon-nucleon and
three-nucleon forces, like the ones derived recently through effective 
field theory and based on chiral symmetry \cite{epelbaum}.

Nevertheless some approaches already exist which directly attack the
6-nucleon problem and beyond 
 with realistic nuclear forces, namely in the method of 
no-core shell model (NCSM) \cite{ncsm} and the Greens function
\cite{gfmc} Monte Carlo treatment.
 In \cite{ncsm} chiral two-nucleon and three-nucleon forces were used
 and applied to
 $^7Li$ with some under binding. In \cite{gfmc} the AV18 
 nucleon-nucleon interaction  and a Urbana three-nucleon
   force was used 
again leading to  some under binding, now for $^6He$. 
 There is also the stochastic variational 
   Monte Carlo method \cite{varga} which, however, still applied
   simplified forces. 

The exploration of chiral
 forces goes on, also including explicitely the $\Delta$-degree 
of freedom \cite{evgeny}, which calls for an increased effort to
establish rigorous approaches beyond $A=4$.  
 The achievments in \cite{ncsm} and \cite{gfmc} 
 demonstrate that a direct treatment of 6 nucleons is feasible on
present day computers and therefore we felt that  another approach, 
the exact formulation within the Yakubovsky equations, is timely. 
About 20 years ago an analogous step turned out to be very fruitful, 
namely the exact formulation of the $\alpha$-particle within 
the Yakubovsky scheme \cite{kamada1,kamada2}. This pioneering
study opened the way to a nowadays 
standard treatment \cite{kamada3,kamada4,nogga,nogga1}
     and allows the inclusion of the most modern two- and three-nucleon
     forces and even first estimates of four-nucleon forces \cite{nogga2}.

In section \ref{sec_two} we apply the 
Yakubovsky equations \cite{yakubovsky}  to 
the six-body problem using the basic notation for sub clusters \cite{mybook}.
        In section \ref{sec_three} we add 
the identity of the nucleons which leads 
to a set of 5 coupled Yakubovsky equations related to 5 different 
sequential sub clusterings of 6 particles.
 In view of the expectation for the dominant  structure of $^6He$,
namely an $\alpha$-core and two loosely bound neutrons, we stop the sequential 
sub clustering with 3 fragments, though the additional step with 
two fragments could be easily performed.

In section \ref{sec_four} and the Appendices we provide technicalities which we
consider  useful for a numerical performance. 
Finally we summarize in section \ref{sec_five}.

\section{The Yakubovsky approach to 6 particles}
\label{sec_two}

We use the standard notation $ a_n $ to denote the various members of
n fragments for a total of N particles. Here N=6 and $ a_2, a_3, a_4,
a_5 $ denote two-, three-,  up to 5-body fragmentations. 5-body
fragmentations necessarily have one pair left and thus $ a_5 $ can be
used to point to a specific pair.

$ a_3 \subset a_2 $ means that the three-body fragments $ a_3 $
consist of  sub clusters out of the two  fragments in 
$ a_2 $ or $ a_3  \supset a_4 $ means that the sub clusters $ a_3 $
when broken up lead to the sub clusters  $ a_4 $.

The bound state $ \Psi$ obeys the homogeneous equation
\begin{eqnarray}
\Psi = G_0 \sum_{ a_5} V_{a_5} \Psi
\label{tot}
\end{eqnarray}
where $ G_0 $ is the 6-particle free Greens operator and $ V_{a_5} $
is a pair force.  The first step is the summation of each pair force to
infinite order. Defining
  \begin{eqnarray}
  \psi_{a_5} \equiv G_0 V_{a_5} \Psi\label{fad}
  \end{eqnarray}
one obtains like for 3 particles
\begin{eqnarray}
\psi_{a_5}  =   G_0 t_{a_5} \sum_{b_5} \overline{\delta_{ a_5 b_5}} \psi_{b_5}
\end{eqnarray}
where $ t_{a_5}$ is a two-body t-operator obeying the Lippmann
Schwinger equation
\begin{eqnarray}
t_{a_5} = V_{a_5} + V_{a_5} G_0 t_{a_5}
\end{eqnarray}
and $ \overline{\delta_{ a_5 b_5}} \equiv 1 - \delta_{ a_5 b_5} $.

Next one defines  new components
\begin{eqnarray}
\psi_{a_5 a_4}  \equiv   G_0 t_{a_5} \sum_{ b_5 \subset a_4} 
\overline{\delta_{ a_5 b_5}} \psi_{b_5}\label{5}
\end{eqnarray}
where all pairs $ a_5, b_5$ are sub clusters of the fragments in $ a_4
$. Clearly
\begin{eqnarray}
\psi_{a_5}  =   \sum_{ a_4 \supset a_5} \psi_{ a_5 a_4}\label{6}
\end{eqnarray}

That relation (\ref{6})  is used to obtain a closed set of equations 
for $ \psi_{a_5 a_4} $:
\begin{eqnarray}
\psi_{a_5 a_4}  =   G_0 t_{a_5} \sum_{ b_5 \subset a_4} 
\overline{\delta_{ a_5 b_5}} \sum_{ b_4 \supset b_5} \psi_{b_5 b_4}
\end{eqnarray}

One separates now the components $  \psi_{a_5 a_4} $ for a given $ a_4
$ from the rest :
  \begin{eqnarray}
& & \psi_{a_5 a_4} - G_0 t_{a_5} \sum_{ b_5 \subset a_4} 
\overline{\delta_{ a_5 b_5}} \psi_{b_5 a_4}
 =   G_0 t_{a_5} \sum_{ b_5 \subset a_4} \overline{\delta_{ a_5
    b_5}} 
\sum_{ b_4 \supset b_5} \overline{\delta_{ a_4 b_4}} \psi_{b_5 b_4}
\label{7}
\end{eqnarray}

Let us define for a fixed $ a_4 $ the column vectors $ \psi^{a_4} $  
and $\psi^{(a_4)} $ with the components
\begin{eqnarray}
 (\psi^{a_4})_{a_5} \equiv  \psi_{a_5 a_4}\label{comp}
 \end{eqnarray}
and
\begin{eqnarray}
 (\psi^{(a_4)})_{b_5} \equiv  \sum_{ b_4 \supset b_5} 
\overline{\delta_{ a_4 b_4}} \psi_{b_5 b_4}\label{comp1}
 \end{eqnarray}

Then  introducing the matrix $ C^{ a_4} $ with the elements 
$ C^{a_4}_{ a_5 b_5} \equiv t_{a_5} \overline{\delta_{ a_5 b_5}} $  
Eq.(\ref{7}) reads
  \begin{eqnarray}
( 1- G_0 C^{a_4}) \psi^{a_4} = G_0 C^{a_4} \psi^{(a_4)}
\end{eqnarray}
 or
  \begin{eqnarray}
 \psi^{a_4} & = & ( 1- G_0 C^{a_4})^{-1} G_0 C^{a_4}  \psi^{(a_4)}
 \equiv G_0 T^{a_4} \psi^{(a_4)}\label{9}
 \end{eqnarray}
Apparently  $ T^{a_4} $ obeys
\begin{eqnarray}
 T^{a_4}   =   C^{a_4} + C^{a_4} G_0 T^{a_4}\label{10}
\end{eqnarray}

In explicite notation (\ref{9}) and (\ref{10}) read
\begin{eqnarray}
\psi_{a_5 a_4}  =   G_0 \sum_{ b_5 \subset a_4} T^{a_4}_{a_5 b_5}  
\psi^{(a_4)}_{b_5} 
=   G_0 \sum_{ b_5 \subset a_4} T^{a_4}_{a_5 b_5} 
 \sum_{ b_4 \supset b_5} 
\overline{\delta_{ a_4 b_4}} \psi_{b_5 b_4}
\label{11}
\end{eqnarray}
\begin{eqnarray}
T^{a_4}_{a_5 b_5}  =   t_{a_5} \overline{\delta_{ a_5 b_5}} +\sum_{
  c_5 \subset a_4}  
t_{a_5} \overline{\delta_{ a_5 c_5}} G_0 T^{a_4}_{c_5 b_5}\label{15}
\end{eqnarray}

Note, there are two types of T-matrices. For $ a_4$ of the type 
$ 123,4,5,6$ $ T_{ a_5 b_5}^{a_4}$ is a $ 3 \times 3 $ matrix and for
$ a_4$ of the type $ 12,34;5,6$  $ T_{ a_5 b_5}^{a_4}$ is a $ 2 \times 2 $ matrix.

Next we further decompose the right hand side of (\ref{11})  according
to  3-body  fragments $a_3$:
\begin{eqnarray}
 \psi_{a_5 a_4}^{ a_3} \equiv  \sum_{ b_5 \subset a_4} G_0 T_{a_5
   b_5}^{a_4} 
\sum_{ b_4 \supset b_5, b_4 \subset a_3} \overline{\delta_{ a_4 b_4}} 
\psi_{b_5 b_4} \label{16}
 \end{eqnarray}
and again
\begin{eqnarray}
 \psi_{a_5 a_4}  =\sum_{a_3 \supset a_4} \psi_{a_5 a_4}^{ a_3}
 \label{14}\end{eqnarray}
is an obvious consequence.

Using again (\ref{14}) Eq. (\ref{16}) can be rewritten as
\begin{eqnarray}
 \psi_{a_5 a_4}^{ a_3}  = \sum_{ b_5 \subset a_4} G_0 T_{a_5
   b_5}^{a_4} 
\sum_{ b_4 \supset b_5, b_4 \subset a_3} \overline{\delta_{ a_4 b_4}}
  \sum_{b_3 \supset b_4} \psi_{b_5 b_4}^{ b_3}
\end{eqnarray}

Analogous to (\ref{7}) one separates $ b_3 = a_3 $ from $ b_3 \ne a_3 $ and gets
\begin{eqnarray}
 & & \psi_{a_5 a_4}^{ a_3}  - G_0 \sum_{ b_5 \subset a_4} T_{a_5
    b_5}^{a_4} 
\sum_{ b_4 \supset b_5, b_4 \subset a_3} \overline{\delta_{ a_4 b_4}}
 \psi_{b_5 b_4}^{ a_3}
  =  G_0 \sum_{ b_5 \subset a_4} T_{a_5 b_5}^{a_4} \sum_{ b_4
   \supset b_5, b_4 
\subset a_3} \overline{\delta_{ a_4 b_4}}
 \psi_{b_5 b_4}^{ (a_3)}
 \label{19}\end{eqnarray}
where we  defined
   \begin{eqnarray}
   \psi_{b_5 b_4}^{ (a_3)} \equiv \sum_{ b_3 \supset b_4} 
\overline{\delta_{ a_3 b_3}}\psi_{b_5 b_4}^{ b_3}
   \end{eqnarray}

Using a matrix notation it is easily seen that (\ref{19}) leads to
\begin{eqnarray}
\psi^{a_3}_{a_5 a_4}  =  G_0 \sum_{ b_4  \subset a_3} 
\sum_{ b_5 \subset b_4}D^{a_3}_{a_5,a_4;b_5 b_4}\psi^{(a_3)}_{b_5 b_4}\label{125}
  \end{eqnarray}
where $ D^{a_3}_{a_5,a_4;b_5 b_4} $ obeys
\begin{eqnarray}
   D^{a_3}_{a_5,a_4;b_5 b_4}  =    T^{a_4}_{a_5 b_5}
   \overline{\delta_{ a_4 b_4}} 
+ \sum_{ c_4  \subset a_3} \sum_{ c_5 \subset a_4}
    T^{a_4}_{a_5 c_5} \overline{\delta_{ a_4 c_4}} G_0 D^{a_3}_{c_5,c_4;b_5 b_4}
   \end{eqnarray}

Note for $ a_3 = 1234,5,6 $ there are 18 pairs of $ a_5,a_4$,  
for $ a_3 = 123,45,6$ there are 9 pairs of $ a_5,a_4$ and for $ a_3 = 12,34,56$
there are 6 pairs of $ a_5,a_4$. This defines the dimensions of 
the different D-matrices. In the following the single nucleons in
$a_n$ will no longer be displayed.

\section{ Implementation of the identity of the nucleons}
\label{sec_three}

We start from (\ref{11}) choosing the case $ a_5 =12$ and $ a_4 =12,34$  
 and obtain 
\begin{eqnarray}
\psi_{12; 12,34} = G_0  T_{12, 12}^{12,34} \psi^{(12,34)}_{12} + 
G_0  T_{12, 34}^{12,34} \psi^{(12,34)}_{34}\label{26}
\end{eqnarray}
which according to (\ref{comp1}) is
\begin{eqnarray}
\psi_{12; 12,34} & = &  G_0  T_{12, 12}^{12,34}  ( \psi_{12,123} +
\psi_{12,124} 
+ \psi_{12,125} + \psi_{12,126}\cr
& + &  \psi_{12; 12,35} + \psi_{12; 12,36} + \psi_{12; 12,45}+
\psi_{12; 12,46}
+ \psi_{12; 12,56})\cr
& + & G_0  T_{12, 34}^{12,34} ( \psi_{34,134} + \psi_{34,234} +
\psi_{34,345} 
+ \psi_{34,346}\cr
& + &  \psi_{34; 34,15} + \psi_{34; 34,16} + \psi_{34; 34,25}+
\psi_{34; 34,26}
+ \psi_{34; 34,56})\label{27}
\end{eqnarray}

It is easily seen, going back to the definitions (\ref{fad}) and
(\ref{5}) together with the antisymmetry requirement for the total 
state $ \Psi$ that
\begin{eqnarray}
  & &    \psi_{34,134} + \psi_{34,234} + \psi_{34,345} + \psi_{34,346}
 +   \psi_{34; 34,15} + \psi_{34; 34,16} + \psi_{34; 34,25}+
  \psi_{34; 34,26}
+ \psi_{34; 34,56}\cr
& =&   P_{13}P_{24} (  \psi_{12,123} + \psi_{12,124} + \psi_{12,125} 
+ \psi_{12,126}\cr
& + &  \psi_{12; 12,35} + \psi_{12; 12,36} + \psi_{12; 12,45}+
\psi_{12; 12,46}
+ \psi_{12; 12,56})\label{28}
\end{eqnarray}

Therefore (\ref{27}) turns into
\begin{eqnarray}
& & \psi_{12; 12,34} = G_0 (  T_{12, 12}^{12,34} +  T_{12, 34}^{12,34}  
P_{13}P_{24})  ( \psi_{12,123} + \psi_{12,124} + \psi_{12,125} + \psi_{12,126}\cr
& + &  \psi_{12; 12,35} + \psi_{12; 12,36} + \psi_{12; 12,45}+
\psi_{12; 12,46}
+ \psi_{12; 12,56})\label{29}
\end{eqnarray}

The coupled equations (\ref{15}) written out for $ a_4 = 12,34$ 
and $ a_5 $ or $ b_5 $ equal to $12$ or $34$  yield when acting with 
$ \tilde P \equiv P_{13}P_{24}$ from both sides 
\begin{eqnarray}
  \tilde P T_{12, 12}^{12,34} \tilde P & = &  T_{34,34}^{12,34}\cr
  \tilde P  T_{12, 34}^{12,34} \tilde P & = &   T_{34,12}^{12,34}
  \end{eqnarray}

Then defining
\begin{eqnarray}
T^{12,34} \equiv  T_{12, 12}^{12,34} + T_{12, 34}^{12,34} \tilde P\label{3.6}
\end{eqnarray}
it follows that $  T^{12,34} $ obeys
\begin{eqnarray}
T^{12,34}   =   t_{12}  \tilde P  + t_{12} G_0 \tilde P T^{12,34}\label{141}
\end{eqnarray}

Therefore (\ref{29}) simplifies to
\begin{eqnarray}
& & \psi_{12; 12,34} = G_0   T^{12,34}  ( \psi_{12,123} +
  \psi_{12,124} 
+ \psi_{12,125} + \psi_{12,126}\cr
& + &  \psi_{12; 12,35} + \psi_{12; 12,36} + \psi_{12; 12,45}+
\psi_{12; 12,46}
+ \psi_{12; 12,56})\label{33}
\end{eqnarray}

Starting again from (\ref{11}) but now for $ a_5 =12$ and $ a_4 = 123$
one obtains
\begin{eqnarray}
& & \psi_{12,123} = G_0 T^{123}_{12,12}  \psi^{(123)}_{12} + G_0
  T^{123}_{12,23}  
\psi^{(123)}_{23}
+ G_0 T^{123}_{12,31}  \psi^{(123)}_{31}\cr
& =&    G_0  T^{123}_{12,12}  ( \psi_{ 12,124} + \psi_{12,125} + \psi_{12,126}\cr
  &+&   \psi_{ 12; 12,34} + \psi_{ 12; 12,35} + \psi_{ 12; 12,36}
 +   \psi_{ 12; 12,45} + \psi_{ 12; 12,46} + \psi_{ 12; 12,56})\cr
& + & G_0 T^{123}_{12,23} ( \psi_{ 23,234} + \psi_{23,235} + \psi_{23,236}\cr
&  + &  \psi_{ 23; 23,14} + \psi_{ 23; 23,15} + \psi_{ 23; 23,16}
 +   \psi_{ 23; 23,45} + \psi_{ 23; 23,46} + \psi_{ 23; 23,56})\cr
& + & G_0 T^{123}_{12,31} ( \psi_{ 31,314} + \psi_{31,315} + \psi_{31,316}\cr
&  + &  \psi_{ 31; 31,24} + \psi_{ 31; 31,25} + \psi_{ 31; 31,26}
 +   \psi_{ 31; 31,45} + \psi_{ 31; 31,46} + \psi_{ 31; 31,56})\cr
& =&    G_0 ( T^{123}_{12,12} +  T^{123}_{12,23} P_{12}P_{23} +
T^{123}_{12,31} 
P_{13}P_{23})
   ( \psi_{ 12,124} + \psi_{12,125} + \psi_{12,126}\cr
&  + &  \psi_{ 12; 12,34} + \psi_{ 12; 12,35} + \psi_{ 12; 12,36}
 +   \psi_{ 12; 12,45} + \psi_{ 12; 12,46} + \psi_{ 12; 12,56})\label{34}
\end{eqnarray}
where we used permutation properties similar as in (\ref{28}).

The corresponding coupled sets (\ref{15}) for $a_4=123$ and using 
relations like $P_{13}P_{23}t_{23}P_{23}P_{13}=t_{12}$ reveals that
\begin{eqnarray}
T^{123} \equiv T^{123}_{12,12} +  T^{123}_{12,23} P_{12}P_{23} 
+ T^{123}_{12,31} P_{13}P_{23}\label{145}
\end{eqnarray}
obeys the equation
\begin{eqnarray}
T^{123} = t_{12} P +  t_{12} P G_0 T^{123}\label{36}
\end{eqnarray}
where $ P \equiv P_{12}P_{23} + P_{13} P_{23} $.

Then (\ref{34}) simplifies to
\begin{eqnarray}
& & \psi_{12,123}  =    G_0  T^{123}  ( \psi_{ 12,124} + \psi_{12,125} 
+ \psi_{12,126}\cr
&  + &  \psi_{ 12; 12,34} + \psi_{ 12; 12,35} + \psi_{ 12; 12,36}
 +   \psi_{ 12; 12,45} + \psi_{ 12; 12,46} + \psi_{ 12; 12,56})\label{3.12}
\end{eqnarray}

The next step is to  decompose $ \psi_{12,123}$  according to
(\ref{16}). For $a_5 =12,a_4 = 123$ the possible  $a_3$'s  are:
  $ 1234 - 1235 - 1236 - 123,45 - 123,46 - 123,56$.

Lets  begin with
\begin{eqnarray}
  & & \psi_{12,123}^{ 1234} \equiv G_0 T_{12, 12}^{123} (
  \psi_{12,124} 
+ \psi_{12; 12,34})\cr
  & + &  G_0 T_{12, 23}^{123} ( \psi_{23,234} + \psi_{23; 23,14})
   +  G_0 T_{12, 31}^{123} ( \psi_{31,134} + \psi_{31; 31,24})
\label{3.13}
  \end{eqnarray}

Since
  \begin{eqnarray}
  \psi_{23,234} + \psi_{23; 23,14} = P_{12}P_{23} ( \psi_{12,124} 
+ \psi_{12; 12,34})\cr
  \psi_{31,134} + \psi_{31; 31,24} = P_{13}P_{23} ( \psi_{12,124} 
+ \psi_{12; 12,34})
  \end{eqnarray}
(\ref{3.13}) simplifies according to (\ref{36}) to
   \begin{eqnarray}
   \psi_{12,123}^{ 1234} = G_0 T^{123} ( \psi_{12,124} + \psi_{12; 12,34})\label{40}
  \end{eqnarray}

Similarily
  \begin{eqnarray}
   \psi_{12,123}^{ 1235} & = &  G_0 T^{123} ( \psi_{12,125} + \psi_{12; 12,35})\cr
   \psi_{12,123}^{ 1236} & = &  G_0 T^{123} ( \psi_{12,126} + \psi_{12; 12,36})
  \end{eqnarray}

Again  using symmetry properties one gets
\begin{eqnarray}
  & & \psi_{12,123}^{ 123,45} = G_0 T^{123}  \psi_{12;12,45}\label{42}
  \end{eqnarray}
  \begin{eqnarray}
  & & \psi_{12,123}^{ 123,46} = G_0 T^{123}  \psi_{12;12,46}
\end{eqnarray}
  \begin{eqnarray}
  & & \psi_{12,123}^{ 123,56} = G_0 T^{123}  \psi_{12;12,56}
  \end{eqnarray}
All summed up
\begin{eqnarray}
& & \psi_{12,123} = \psi_{12,123}^{ 1234} + \psi_{12,123}^{ 1235} 
+ \psi_{12,123}^{ 1236} + \psi_{12,123}^{ 123,45}
 +   \psi_{12,123}^{ 123,46} + \psi_{12,123}^{ 123,56}\label{44}
\end{eqnarray}
agrees with (\ref{3.12}).

Next we decompose (\ref{11}) according to (\ref{16}) for $a_5 =12,a_4=12,34$. 
The possible $a_3$'s are: 
    $1234 - 125,34 - 126,34 - 12,345 - 12,346 - 12,34,56$,
which are now regarded in turn.
\begin{eqnarray}
    & & \psi_{12;12,34}^{1234} = G_0  T_{12, 12}^{12,34} (
  \psi_{12,123} 
+ \psi_{12,124})
      +   G_0  T_{12, 34}^{12,34} ( \psi_{34,234} + \psi_{34,134})
    \end{eqnarray}

Since
    \begin{eqnarray}
     P_{13}P_{24} ( \psi_{12,123} + \psi_{12,124}) = \psi_{34,234} + \psi_{34,134}
    \end{eqnarray}
one can use (\ref{3.6}) and gets 
  \begin{eqnarray}
    \psi_{12;12,34}^{1234} & = &  G_0 ( T_{12, 12}^{12,34} +  T_{12,34}^{12,34}   
P_{13}P_{24})  ( \psi_{12,123} + \psi_{12,124})\cr
    & = & G_0  T^{12,34}  ( \psi_{12,123} + \psi_{12,124})\label{47}
    \end{eqnarray}

Next
    \begin{eqnarray}
    \psi_{12;12,34}^{125,34} & = &  G_0  T_{12,
      12}^{12,34}\psi_{12,125} 
+ G_0   T_{12, 34}^{12,34} ( \psi_{34; 15,34} + \psi_{34;25,34})\cr
    \psi_{12;12,34}^{ 345,12} & = &  G_0  T_{12, 12}^{12,34}(\psi_{12;
      12,35} 
+ \psi_{12;12,45})+ G_0   T_{12, 34}^{12,34}  \psi_{34,345}
    \end{eqnarray}

The two amplitudes $\psi_{12;12,34}^{125,34}$ and $
\psi_{12;12,34}^{345,12}$ can not be related by permutations, but
their sum can be used
\begin{eqnarray}
     \psi_{12;12,34}^{125,34} +  \psi_{12;12,34}^{ 345,12} &=& 
G_0  T_{12, 12}^{12,34}( \psi_{12,125} + \psi_{12; 12,35} + \psi_{12;12,45})\cr
    & + & G_0   T_{12, 34}^{12,34} ( \psi_{34; 15,34} +
\psi_{34;25,34}+ \psi_{34,345})
    \end{eqnarray}
in the sense
\begin{eqnarray}
   P_{13}P_{24}     ( \psi_{12,125} + \psi_{12; 12,35} +
   \psi_{12;12,45}) 
= \psi_{34; 15,34} + \psi_{34;25,34}+ \psi_{34,345}
\end{eqnarray}

This leads to
\begin{eqnarray}
   \psi_{12;12,34}^{125,34} +  \psi_{12;12,34}^{ 345,12} = G_0
   T^{12,34}( \psi_{12,125}  + \psi_{12; 12,35} + \psi_{12;12,45})\label{51}
    \end{eqnarray}

Similarily
\begin{eqnarray}
   \psi_{12;12,34}^{126,34} + \psi_{12;12,34}^{346,12} 
=  G_0  T^{12,34}( \psi_{12,126}  + \psi_{12; 12,36} + \psi_{12;12,46})\label{52}
    \end{eqnarray}
and finally
\begin{eqnarray}
    \psi_{12;12,34}^{12,34,56} = G_0  T_{12, 12}^{12,34} \psi_{12;
      12,56} 
+ G_0  T_{12, 34}^{12,34} \psi_{34; 34,56} = G_0  T^{12,34}  
\psi_{12; 12,56}\label{53}
    \end{eqnarray}

Thus, Eqs.(\ref{47}), (\ref{51})-(\ref{53}),    summarizes   to
 \begin{eqnarray}
    & & \psi_{12;12,34} = \psi_{12;12,34}^{1234} +
   \psi_{12;12,34}^{125,34}
+ \psi_{12;12,34}^{345,12} 
     +    \psi_{12;12,34}^{126,34} + \psi_{12;12,34}^{346,12}+ 
\psi_{12;12,34}^{12,34,56}\label{54}
\end{eqnarray}
which when written out agrees with (\ref{33}).

The two amplitudes $ \psi_{12,123}^{1234}$ and $\psi_{12;12,34}^{1234}$ 
expressed in (\ref{40}) and (\ref{47}) are connected to each other as shown now.
 The expression (\ref{44}) can easily be converted to $ \psi_{12,124}$
and using in addition (\ref{54}) one finds
\begin{eqnarray}
& & \psi_{12,123}^{ 1234} - G_0 T^{123} ( \psi_{12,124}^{ 1234} 
+ \psi_{12;12,34}^{ 1234})\cr
 & = &  G_0 T^{123} ( \psi_{12,124}^{ 1245} + \psi_{12,124}^{ 1246} 
+ \psi_{12,124}^{ 124,35}
 +  \psi_{12,124}^{ 124,36 } + \psi_{12,124}^{ 124,56 } 
+  \psi_{12;12,34}^{125,34 }\cr
 & + &  \psi_{12;12,34}^{345,12 } +  \psi_{12;12,34}^{126,34 }
+ \psi_{12;12,34}^{346,12 }
   +   \psi_{12;12,34}^{ 12,34, 56 })\label{55}
\end{eqnarray}

Correspondingly  (\ref{47})  yields
\begin{eqnarray}
& & \psi_{12;12,34}^{ 1234} - G_0  T^{12,34} (  \psi_{12,123}^{ 1234} 
+\psi_{12,124}^{ 1234}) \cr
 & = &  G_0  T^{12,34}  ( \psi_{12,123}^{ 1235} +  \psi_{12,123}^{
  123,45 } 
+ \psi_{12,123}^{ 1236}
  +   \psi_{12,123}^{123,46} +   \psi_{12,123}^{123,56}  
+   \psi_{12,124}^{ 1245}\cr
 &  + &   \psi_{12,124}^{ 124,35 } + \psi_{12,124}^{ 1246}  
+   \psi_{12,124}^{124,36} + \psi_{12,124}^{124,56})
\label{56}
\end{eqnarray}

With $ \psi_{12,124}^{ 1234}  = - P_{34} \psi_{12,123}^{ 1234}$  we 
can put (\ref{55}) and (\ref{56}) into a matrix form:
\begin{eqnarray}
    & & \left(
   \begin{array}{rcl}
   \psi_{12,123}^{ 1234}\\
   \psi_{12;12,34}^{ 1234}\\
   \end{array}
   \right) - G_0 \left(
   \begin{array}{rcl}
   -T^{123} P_{34}  & T^{123} \\
    T^{12,34} ( 1-P_{34}) & 0 \\
   \end{array}
   \right)
   \left(
   \begin{array}{rcl}
   \psi_{12,123}^{ 1234}\\
   \psi_{12;12,34}^{ 1234}\\
   \end{array}
   \right)
   \cr
   & = &    G_0  \left(
   \begin{array}{rcl}
    T^{123} &(&  \psi_{12,124}^{ 1245} + \psi_{12,124}^{ 1246} 
+ \psi_{12,124}^{ 124,35}
 +  \psi_{12,124}^{ 124,36 } + \psi_{12,124}^{ 124,56 } 
+   \psi_{12;12,34}^{125,34 }\cr
 & + &  \psi_{12;12,34}^{345,12 } +  \psi_{12;12,34}^{126,34 }
+ \psi_{12;12,34}^{346,12 } 
  +    \psi_{12;12,34} ^{ 12,34, 56 })\\
    T^{12,34} &(&  \psi_{12,123}^{ 1235} +  \psi_{12,123}^{ 123,45 } 
+ \psi_{12,123}^{ 1236} + \psi_{12,123}^{123,46}
  +   \psi_{12,123}^{123,56}  +   \psi_{12,124}^{ 1245} \cr
&+&  \psi_{12,124}^{ 124,35 } + \psi_{12,124}^{ 1246} 
  +   \psi_{12,124}^{124,36} + \psi_{12,124}^{124,56})\\
   \end{array}
   \right)
\label{57}
\end{eqnarray}

Since
\begin{eqnarray}
   \psi_{12,124} ^{1245} & = &  - P_{34} \psi_{12,123}^{1235}\cr
   \psi_{12,124} ^{124, 35} & = &  - P_{34} \psi_{12,123}^{123, 45}
   \end{eqnarray}
the right hand side of (\ref{57}) can be  factored and (\ref{57}) 
achieves the form
\begin{eqnarray}
 & &    \left(
   \begin{array}{rcl}
   \psi_{12,123}^{ 1234}\\
   \psi_{12;12,34}^{ 1234}\\
   \end{array}
   \right) - G_0 \left(
   \begin{array}{rcl}
   -T^{123} P_{34}  & T^{123} \\
    T^{12,34} ( 1-P_{34}) & 0 \\
   \end{array}
   \right)
   \left(
   \begin{array}{rcl}
   \psi_{12,123}^{ 1234}\\
   \psi_{12;12,34}^{ 1234}\\
   \end{array}
   \right)\cr
  & = &  G_0 \left(
   \begin{array}{rcl}
   T^{123}(-P_{34})  & T^{123} \\
    T^{12,34} ( 1 - P_{34})  & 0\\
   \end{array}
   \right)
   \left(
   \begin{array}{rcl}
    & & \psi_{12,123}^{ 1235} + \psi_{12,123}^{ 123,45 }+ \psi_{12,123}^{ 1236}\cr
      &+&   \psi_{12,123}^{ 123,46 } + \psi_{12,123}^{ 123,56 }  \\
    & &  \psi_{12;12,34}^{125,34 } + \psi_{12;12,34}^{345,12 } 
+   \psi_{12;12,34}^{126,34 }\cr
     &+&   \psi_{12;12,34}^{346,12 } +  \psi_{12;12,34}^{12,34,56 }\\
   \end{array}
   \right)
\end{eqnarray}

The right hand side can be reduced applying permutations and one obtains
\begin{eqnarray}
 & &    \left(
   \begin{array}{rcl}
   \psi_{12,123}^{ 1234}\\
   \psi_{12;12,34}^{ 1234}\\
   \end{array}
   \right) - G_0 \left(
   \begin{array}{rcl}
   -T^{123} P_{34}  & T^{123} \\
    T^{12,34} ( 1-P_{34}) & 0 \\
   \end{array}
   \right)
   \left(
   \begin{array}{rcl}
   \psi_{12,123}^{ 1234}\\
   \psi_{12;12,34}^{ 1234}\\
   \end{array}
   \right)\cr
  & = &  G_0 \left(
   \begin{array}{rcl}
   T^{123}(-P_{34})  & T^{123} \\
    T^{12,34} ( 1 - P_{34})  & 0\\
   \end{array}
   \right)
         \left(
   \begin{array}{rcl}
    - ( P_{45} + P_{46}) \psi_{ 12,123}^{1234} 
+ ( 1 - P_{56} - P_{46}) \psi_{12,123}^{ 123,45 }\\
     (1 - P_{56}) (\psi_{12;12,34}^{125,34 } + \psi_{12;12,34}^{345,12
}) +  \psi_{12;12,34}^{12,34,56 }\\
   \end{array}
   \right)
   \label{60}
   \end{eqnarray}
which can be put into the form     $\left(
   \begin{array}{rcl}
   \psi_{12,123}^{ 1234}\\
   \psi_{12;12,34}^{ 1234}\\
   \end{array}
   \right) $ with the result 
\begin{eqnarray}
   \left(
   \begin{array}{rcl}
   \psi_{12,123}^{ 1234}\\
   \psi_{12;12,34}^{ 1234}\\
   \end{array}
   \right)
   \equiv G_0 \left(
   \begin{array}{rcl}
   D_{11} & D_{12} \\
   D_{21} & D_{22}\\
   \end{array}
   \right)
  \left(
   \begin{array}{rcl}
    - ( P_{45} + P_{46}) \psi_{ 12,123}^{1234} + ( 1 - P_{56} 
- P_{46}) \psi_{12,123}^{ 123,45 }\\
     (1 - P_{56}) (\psi_{12;12,34}^{125,34 } +
 \psi_{12;12,34}^{345,12 }) +  
\psi_{12;12,34}^{12,34,56 }\\
   \end{array}
   \right)
   \label{61}
\end{eqnarray}
where  the matrix $ D$ obeys
\begin{eqnarray}
  & &    \left(
   \begin{array}{rcl}
   D_{11} & D_{12} \\
   D_{21} & D_{22}\\
   \end{array}
   \right) =   \left(
   \begin{array}{rcl}
   -T^{123} P_{34}  & T^{123} \\
    T^{12,34} ( 1-P_{34}) & 0 \\
   \end{array}
   \right)\cr
   &+&  \left(
   \begin{array}{rcl}
   -T^{123} P_{34}  & T^{123} \\
    T^{12,34} ( 1-P_{34}) & 0 \\
   \end{array}
   \right)
   G_0
      \left(
   \begin{array}{rcl}
   D_{11} & D_{12} \\
   D_{21} & D_{22}\\
   \end{array}
   \right)
   \label{62}
\end{eqnarray}

For the numerical treatment, however,  we consider the structure (\ref{60})
to be more advantageous.

The right hand side  of (\ref{60}) contains new amplitudes. 
After adequate permutation of (\ref{54}) one obtains from (\ref{42})
\begin{eqnarray}
\psi_{12,123}^{ 123,45}  & = & G_0 T^{123}  (\psi_{12;12,45}^{ 1245} 
+ \psi_{12;12,45}^{ 123,45} + \psi_{12;12,45}^{ 345,12}
  +   \psi_{12;12,45}^{ 456,12} + \psi_{12;12,45}^{ 126,45} 
+ \psi_{12;12,45}^{ 12,45,36})
\end{eqnarray}
or
\begin{eqnarray}
& & \psi_{12,123}^{ 123,45} - G_0 T^{123} (-( 1- P_{36}) P_{53} (  
\psi_{12;12,34}^{ 125,34} + \psi_{12;12,34}^{ 345,12}))\cr
 &=&  G_0 T^{123} (  - P_{35} \psi_{12;12,34}^{ 1234}  
+ \psi_{12;12,45}^{ 12,45,36})
\end{eqnarray}

Further  (\ref{51}) yields inserting the decomposition of the right 
hand side related to (\ref{44}) and (\ref{54})  yields
\begin{eqnarray}
& & \psi_{12;12,34}^{ 125,34} + \psi_{12,12,34}^{ 345,12}  
= G_0  T^{12,34}   (\psi_{12,125}^{1235} + \psi_{12,125}^{1245} 
+ \psi_{12,125}^{125,34}\cr
 & + &  \psi_{12,125}^{1256} + \psi_{12,125}^{125,36}+  \psi_{12,125}^{125,46} 
  +  \psi_{12;12,35}^{ 1235} + \psi_{12;12,35}^{ 124,35} 
+ \psi_{12;12,35}^{ 345,12}\cr
&  +  & \psi_{12;12,35}^{ 356,12} + \psi_{12;12,35}^{ 126,35} 
+ \psi_{12;12,35}^{ 12,56,34}
 +  \psi_{12;12,45}^{ 1245} + \psi_{12;12,45}^{ 123,45} 
+ \psi_{12;12,45}^{ 345,12}\cr
&  +  & \psi_{12;12,45}^{ 456,12} + \psi_{12;12,45}^{ 126,45} 
+ \psi_{12;12,45}^{ 12,34,56}
\end{eqnarray}

Here quite a few amplitudes can be related to previous ones by permutations 
leading to:
\begin{eqnarray}
& & \psi_{12;12,34}^{ 125,34} + \psi_{12,12,34}^{ 345,12}  -  G_0  T^{12,34}
 (  P_{35} P_{56} - P_{35} - P_{46} - P_{45}) ( \psi_{12;12,34}^{
    125,34} + \psi_{12;12,34}^{ 345,12})\cr
 &  = &  G_0  T^{12,34} ( ( 1- P_{36} - P_{34} - P_{46})P_{34} P_{35}  
\psi_{12,123}^{1234} + ( 1 - P_{34}) P_{46}P_{35}\psi_{12,123}^{123,45}\cr
&  -& ( P_{45} + P_{35}) \psi_{12;12,34}^{ 1234} - P_{35} 
\psi_{12;12,34}^{ 12,34,56})
\end{eqnarray}

Finally we regard $ \psi_{12;12,34}^{ 12,34,56}$, which after
(\ref{53}) and  suitable permutations of (\ref{54}) turns into
\begin{eqnarray}
& & \psi_{12;12,34}^{ 12,34,56}  - G_0T^{12,34}P_{35} P_{46}  
\psi_{12;12,34}^{ 12,34, 56}\cr
&   = & G_0   T^{12,34} ( P_{35} P_{46} \psi_{12;12,34}^{ 1234} 
+ ( 1-P_{34})P_{35}P_{46}( \psi_{12;12,34}^{ 125,34} +  \psi_{12;12,34}^{ 345,12}))
\end{eqnarray}
where we used
\begin{eqnarray}
\psi_{12;12,56}^{ 12,56,34} & =  & G_0 T^{12,56} \psi_{12;12,34}
 =    P_{35} P_{46} G_0 T^{12,34}  \psi_{12;12,56} =  
P_{35} P_{46} \psi_{12;12,34}^{ 12,34,56} 
\end{eqnarray}

Thus we end up with 5 independent amplitudes:  
$ \psi_{12,123}^{ 1234}$, $\psi_{12;12,34}^{ 1234}$, $\psi_{12,123}^{
  123,45}$, $\psi_{12;12,34}^{ 125,34} + \psi_{12,12,34}^{ 345,12}$,
and $\psi_{12;12,34}^{ 12,34,56}$, coupled in the equations
\begin{eqnarray}
 & &    \left(
   \begin{array}{rcl}
   \psi_{12,123}^{ 1234}\\
   \psi_{12;12,34}^{ 1234}\\
   \end{array}
   \right) - G_0 \left(
   \begin{array}{rcl}
   -T^{123} P_{34}  & T^{123} \\
    T^{12,34} ( 1-P_{34}) & 0 \\
   \end{array}
   \right)
   \left(
   \begin{array}{rcl}
   \psi_{12,123}^{ 1234}\\
   \psi_{12;12,34}^{ 1234}\\
   \end{array}
   \right)\cr
  & = &  G_0 \left(
   \begin{array}{rcl}
   T^{123}(-P_{34})  & T^{123} \\
    T^{12,34} ( 1 - P_{34})  & 0\\
   \end{array}
   \right)
         \left(
   \begin{array}{rcl}
    - ( P_{45} + P_{46}) \psi_{ 12,123}^{1234} + ( 1 - P_{56} -
    P_{46}) \psi_{12,123}^{ 123,45 }\\
     (1 - P_{56}) (\psi_{12;12,34}^{125,34 } + \psi_{12;12,34}^{345,12
    }) +  \psi_{12;12,34}^{12,34,56 }\\
   \end{array}
   \right)
   \label{70}
   \end{eqnarray}

\begin{eqnarray}
& & \psi_{12,123}^{ 123,45} - G_0 T^{123} (-( 1- P_{36}) P_{53} (  
\psi_{12;12,34}^{ 125,34} + \psi_{12;12,34}^{ 345,12}))\cr
& = &  - G_0 T^{123} P_{35}( \psi_{12;12,34}^{ 1234}  
+ \psi_{12;12,34}^{ 12,34,56})\label{71}
\end{eqnarray}
\begin{eqnarray}
& & \psi_{12;12,34}^{ 125,34} + \psi_{12,12,34}^{ 345,12}  -  G_0  T^{12,34}
 (  P_{35} P_{56} - P_{35} - P_{46} - P_{45}) ( \psi_{12;12,34}^{
    125,34} + \psi_{12;12,34}^{ 345,12})\cr
 &  = &  G_0  T^{12,34} ( ( 1- P_{36} - P_{34} - P_{46})P_{34} P_{35}  
\psi_{12,123}^{1234} + ( 1 - P_{34}) P_{46}P_{35}\psi_{12,123}^{123,45}\cr
&  -& ( P_{45} + P_{35}) \psi_{12;12,34}^{ 1234} -P_{35} 
\psi_{12;12,34}^{ 12,34,56})\label{72}
\end{eqnarray}

\begin{eqnarray}
& & \psi_{12;12,34}^{ 12,34,56}  - G_0T^{12,34}P_{35} P_{46}  
\psi_{12;12,34}^{ 12,34, 56}
 =  G_0   T^{12,34} ( P_{35} P_{46} \psi_{12;12,34}^{ 1234}\cr
&  + &  ( 1-P_{34})P_{35}P_{46}( \psi_{12;12,34}^{ 125,34} +  
\psi_{12;12,34}^{ 345,12}))\label{73}
\end{eqnarray}

It remains to establish the expression for the total state $ \Psi $. 
We use (\ref{tot}), (\ref{5}), (\ref{44}), (\ref{54})
 together with permutations and obtain
\begin{eqnarray}
\Psi & = &  [  1 - P_{23} - P_{24} - P_{13} - P_{14}  + P_{13} P_{24}]
[ ( 1 - P_{34}) \psi_{12,123}^{1234} +  \psi_{12;12,34}^{1234}]\cr
& - &  [  1 - P_{23} - P_{24} - P_{13} - P_{14}  + P_{13} P_{24}]\cr
&& [ ( ( 1 - P_{34}) ( P_{45} + P_{46} + P_{35} +  P_{36})
- ( P_{35}P_{46} + P_{36} P_{45})) \psi_{12,123}^{1234}\cr
& + & ( P_{45} + P_{46} + P_{35} +  P_{36} -  P_{35} P_{46}) 
\psi_{12;12,34}^{1234}]\cr
& - & [  P_{25} + P_{26} + P_{15} + P_{16} -    P_{13} P_{25} 
-  P_{13} P_{26} -  P_{14} P_{25} -  P_{14} P_{26} -  P_{15} P_{26}]\cr
& & [ ( 1 - P_{34}) \psi_{12,123}^{1234} +  \psi_{12;12,34}^{1234}\cr
& - & ( (  1 -P_{34}) ( P_{45} + P_{46} + P_{35} +  P_{36})- ( P_{35}
      P_{46} 
+ P_{36} P_{45})) \psi_{12,123}^{1234}\cr
& - & ( P_{45} + P_{46} + P_{35} +  P_{36} -  P_{35} P_{46}) 
\psi_{12;12,34}^{1234}]\cr
& + & [  1 - P_{23} - P_{24} - P_{13} - P_{14}  + P_{13} P_{24}]
 [( 1 - P_{34}  -   P_{35} - P_{36}) (   1 -  P_{56} - P_{46}) 
\psi_{12,123}^{123,45}\cr
& + & (1 -  P_{45} - P_{46} - P_{35} -  P_{36}  + P_{35} P_{46})
 ( ( 1 - P_{56}) ( \psi_{12;12,34}^{125,34} +
\psi_{12;12,34}^{345,12}) 
+ \psi_{12;12,34}^{12,34,56})]\cr
&   - & [  P_{25} + P_{26} + P_{15} + P_{16}   -  P_{13} P_{25} 
-  P_{13} P_{26} -  P_{14} P_{25} -  P_{14} P_{26} -  P_{15} P_{26}]\cr
& & [  ( 1 - P_{34}  -   P_{35} - P_{36}) (   1 -  P_{56} - P_{46}) 
\psi_{12,123}^{123,45}
+ (1 -  P_{45} - P_{46} - P_{35} -  P_{36}  + P_{35} P_{46})\cr
& & ( ( 1 - P_{56}) ( \psi_{12;12,34}^{125,34} +
\psi_{12;12,34}^{345,12}) 
+ \psi_{12;12,34}^{12,34,56})]
\label{TOT}
\end{eqnarray}

The first piece
\begin{eqnarray}
\Psi_4 & \equiv  &  [  1 - P_{23} - P_{24} - P_{13} - P_{14}  + P_{13} P_{24}]
[ ( 1 - P_{34}) \psi_{12,123}^{1234} +  \psi_{12;12,34}^{1234}]
\end{eqnarray}
has exactly the form for a 4-nucleon bound state
\cite{kamada1,kamada3} 
based on the two Yakubovsky components
 $ \psi_{12,123}^{1234}$ and $  \psi_{12;12,34}^{1234} $. Now for 
6 nucleons these two amplitudes depend not only
 on the momenta of nucleons $1$ to $4$ but  also on the  momenta 
of nucleons 5 and 6. In the spirit of the effective 3-body
  model $ \alpha-n-n $ $\Psi_4$ factorizes into a product of the 
$\alpha$-state and the wave function for the two neutrons.
   The next two pieces, still related to the same two Yakubovsky
   components,  
antisymmetrize the nucleons 5 and 6 
    in relation to the nucleons 1 to 4. The remaining two pieces, going 
with the additional three Yakubovsky components
 $\psi_{12,123}^{123,45}$, $ \psi_{12;12,34}^{125,34} +
\psi_{12;12,34}^{345,12}$ 
and $ \psi_{12;12,34}^{12,34,56} $,  
   go clearly beyond the effective 3-body model 
and allow for additional sub clusterings of different  types.
  In addition one has to keep in mind that the most general form (\ref{TOT}) 
  allows many more distributions of neutrons and protons 
  in the sub clusters than restricted in the effective $\alpha-n-n$ picture.

\section{ Technicalities for a numerical implementation}
\label{sec_four}

The 5 independent amplitudes require 5 different Jacobi momenta. 
For $ \psi_{12,34}^{ 123} $
we choose
\begin{eqnarray}
    \vec a_1 & = &  \frac{1}{2} ( \vec k_1 - \vec k_2)\cr
 \vec a_2  & = &  \frac{1}{3} ( 2 \vec k_3 - \vec k_1 - \vec k_2)\cr
   \vec a_3  & = &  \frac{1}{4} ( 3 \vec k_4 - \vec k_1 
- \vec k_2 -\vec k_3 )\cr
  \vec a_4  & = &  \frac{1}{2} ( \vec k_5 - \vec k_6)\cr
    \vec a_5  & = &  \frac{1}{3} ( 2 ( \vec k_5 + \vec k_6)  
- \vec k_1 - \vec k_2 -\vec k_3 - \vec k_4 )\label{74}
\end{eqnarray}

Then  the individual momenta (under the condition $ \sum_i \vec k_i
=0$) in terms of those Jacobi momenta are
\begin{eqnarray}
  \vec k_1 & = &  \vec a_1 -   \frac{1}{2} \vec a_2 - \frac{1}{3}\vec
  a_3 
- \frac{1}{4}\vec a_5\cr
  \vec k_2  & = &  - \vec a_1 -   \frac{1}{2} \vec a_2 -
  \frac{1}{3}\vec a_3 
- \frac{1}{4}\vec a_5\cr
  \vec k_3  & = &  \vec a_2 -  \frac{1}{3}\vec a_3 - \frac{1}{4}\vec a_5\cr
  \vec k_4  & = &  \vec a_3 - \frac{1}{4}\vec a_5\cr
  \vec k_5  & = &  \vec a_4 + \frac{1}{2}\vec a_5
  \end{eqnarray}

The kinetic energy is
\begin{eqnarray}
   \sum_{i=1}^{6}\frac{k_i^2}{2m} = \frac{1}{m} ( a_1^2 + \frac{3}{4}
   a_2^2 
+ \frac{2}{3} a_3^2 + a_4^2 + \frac{3}{8} a_5^2)
  \end{eqnarray}

The remaining choices of Jacobi momenta are given in Appendix \ref{appendix_a}.

In a partial wave decomposition the basis states suitable 
for $ \psi_{12,34}^{ 123} $ are
\begin{eqnarray}
  & &  |a_1 a_2 a_3 a_4 a_5 \alpha_1>\cr
   &\equiv& | a_1 a_2 a_3 a_4 a_5, ( l_1 s_{12}) j_1 
(l_2 \frac{1}{2})j_2 ( j_1 j_2) I_{3} (l_3\frac{1}{2}) j_4 ( I_3 j_4) 
I_4  (l_5 s_{56})j_5 (l_4 j_5)I_{5}
    ( I_{4} I_{5}) J M> \cr
   & & | (t_{12}\frac{1}{2})t_3 ( t_3 \frac{1}{2}) t_4 ( t_4 t_{56}) T
M_T>
\label{77}
  \end{eqnarray}
Here the orbital angular momenta $l_i$ go with the $ \vec a_i$,
$s_{ij} $ are 
two-body spins for particles $ij$, $j_i$ are total 1- and  2-body  angular
   momenta coupled out of orbital and spin angular momenta, $I_{3} $ 
and $ I_4$ are  total 3- and 4-body  angular momenta, $I_5$ the
     total angular momentum of particles 5 and 6 relative 
to particles 1-4 and finally $ I_4$ and $I_5$ are coupled to $J$, the 
conserved total 6-body angular momentum.  The second state in (\ref{77})
   refers to isospin in an obvious manner.

Using (\ref{77}) the amplitude $ \psi^{1234}_{123}$ has the representation
\begin{eqnarray}
  |\psi^{1234}_{123}> = \sum_{\alpha_1}\ \int \Pi_{i=1}^5 da_i a_i^2 
| a_1 a_2 a_3 a_4 a_5 \alpha_1> < a_1 a_2 a_3 a_4 a_5 \alpha_1| \psi^{1234}_{123}>
\label{78} 
\end{eqnarray}
where the discrete set of quantum  numbers $ \alpha_1$ runs   over 
all values for a given $J$ and $T$.

Thereby antisymmetry requires that $ l_1 + s_{12} + t_{12}$  
and $l_4 + s_{56} + t_{56}$ have to be odd.

Analogous basis states for the other 4 amplitudes and related Jacobi 
momenta should be obvious. One other example is given below.

If one projects the coupled equations (\ref{70}) - (\ref{73}) from the
left onto the adequate basis states and expands the 5 amplitudes on the right
       hand side like in (\ref{78}) one faces the task to evaluate 
the various kernels. As an example out of (\ref{70}) we take
\begin{eqnarray}
   & & < a_1 a_2 a_3 a_4 a_5 \alpha_1|\psi^{1234}_{12,123}>\cr
    &=& - G_0  < a_1 a_2 a_3 a_4 a_5 \alpha_1|  T^{123} P_{34}
    \sum_{\alpha_1'} \int \Pi_{i=1}^5 da_i' a_i^{'2} | a_1' a_2' a_3'
    a_4' a_5' 
\alpha_1'> \cr
    & & < a_1' a_2' a_3' a_4' a_5' \alpha_1'| \psi^{1234}_{12,123}>\cr
    &+&  G_0  < a_1 a_2 a_3 a_4 a_5 \alpha_1|  T^{123}|
\sum_{\alpha_2'} 
\int \Pi_{i=1}^5 db_i' b_i^{'2} | b_1' b_2' b_3' b_4' b_5' \alpha_2'>\cr
   & &  < b_1' b_2' b_3' b_4' b_5' \alpha_2'| \psi^{1234}_{12;12,34}> +
    \cdots
\label{79}
\end{eqnarray}
where the $ \cdots$ refer to the remaining operators and amplitudes
and the b-states are related to $ \psi_{12;12,34}^{1234}$.

Using techniques like the ones presented in 
\cite{mybook,kamada1,nogga1,witala} it is straightforward, though tedious,
 to generate the kernels like 
$ < a_1 a_2 a_3 a_4 a_5 \alpha_1|  T^{123} P_{34}
 | a_1' a_2' a_3' a_4' a_5' \alpha_1'> $
       or $ < b_1  b_2 b_3 b_4 b_5 \alpha_2| T^{12,34} ( 1-P_{34}) (
       P_{45}
+ P_{46}) | a_1' a_2' a_3' a_4' a_5' \alpha_1'>$.

In the course of the required recoupling among the different Jacobi 
momenta the variables for the 5 amplitudes on the right hand sides are 
in general linear combinations of intermediate integration variables 
which include angles besides momentum magnitudes. An example
illustrates that situation:
\begin{eqnarray}
  H \equiv P_{45}  \psi_{12,123}^{1234}
  \end{eqnarray}

We project onto the basis states for Jacobi momenta of type b given in 
Appendix \ref{appendix_a}:
\begin{eqnarray}
   | b > & \equiv &  | b_1 b_2 b_3 b_4 b_5; ( l_1 s_{12}) j_1 ( l_2
   s_{34}) j_2 
( j_1 j_2 ) S ( L S ) I ( l_4 s_{56}) j_5 ( l_5 j_5) I_5 ( I I_5 ) JM>\cr
   & & | (t_{12} t_{34}) t_{1-4} ( t_{1-4} t_{56} ) T M_T>\label{81}
\end{eqnarray}
where the orbital angular momenta $l_i, i\ne 3$ go with $\vec b_i$
and 
$L$ goes with $\vec b_3$. The 2-body spins are $s_{12}, s_{34}$ and
$s_{56}$, 
$I$ is the total angular momentum for particles 1-4 and $I_5$ the
total 
angular momentum of the pair 56 against the subsystem 1-4. They are
coupled 
to the total angular momentum $J$, which is conserved. The isospin
coupling 
should be obvious.

For the sake of simplicity we choose  s-waves which simplifies (\ref{81}) to
\begin{eqnarray}
    | b > & \equiv &   \delta_{S I} \delta_{ s_{12} j_1}\delta_{
      s_{34} j_2}
\delta_{ j_5 I_5} \delta_{ s_{56} j_5}\cr
  & & | b_1 b_2 b_3 b_4 b_5>
   | ( s_{12} s_{34} ) S ( S s_{56}) JM> |(t_{12} t_{34}) t_{1-4} 
( t_{1-4} t_{56} ) T M_T>
  \end{eqnarray}

The basis states (\ref{77}) related to the Jacobi momenta of type a 
and restricted to s-waves are
\begin{eqnarray}
| a > & = &  \delta_{ s_{12} j_1} \delta_{ j_2 \frac{1}{2}} 
\delta_{j_4 \frac{1}{2}} \delta_{ j_5 s_{56}} \delta_{ j_5 I_5}
  | a_1 a_2 a_3 a_4 a_5> \cr
&& | ( s_{12} \frac{1}{2}) I_3 
( I_3 \frac{1}{2} ) I_4 ( I_4 s_{56}) JM>
 | (t_{12} \frac{1}{2}) t_3 ( t_3 \frac{1}{2} )t_4 ( t_4 t_{56})  T
M_T>
\label{88}
\end{eqnarray}

Then $ \psi_{12,123}^{1234} $ has the representation
\begin{eqnarray}
   |\psi_{12,123}^{1234}> \equiv \sum_{\alpha_1} \delta \cdots 
\int \Pi_{i=1}^5 da_i a_i^2 | a_1 a_2 a_3 a_4 a_5 > | spin>_a | isospin>_a
    < a|\psi_{12,123}^{1234}>\label{89}
  \end{eqnarray}
and H projected from the left is
\begin{eqnarray}
   < b| H> & = &  \delta_{S I} \delta_{ s_{12} j_1}\delta_{ s_{34}
     j_2}
\delta_{ j_5 I_5} \delta_{ s_{56} j_5}\cr
   & & < (t_{12} t_{34}) t_{1-4} ( t_{1-4} t_{56} ) T M_T|  <( s_{12}
s_{34} ) 
S ( S s_{56}) J M | <  b_1 b_2 b_3 b_4 b_5| P_{45}\cr
   & & \sum_{\alpha_1'} \delta \cdots \int \Pi_{i=1}^5 da_i a_i^2 
| a_1 a_2 a_3 a_4 a_5 >
   | ( s_{12}' \frac{1}{2}) I_3' ( I_3' \frac{1}{2} ) I_4' ( I_4'
   s_{56}') JM>
\cr
& & | (t_{12}' \frac{1}{2}) t_3' ( t_3' \frac{1}{2} )t_4' ( t_4'
t_{56}')  T M_T>
< a|\psi_{12,123}^{1234}>\cr
& \equiv &  \delta \cdots  \sum_{\alpha_1} \delta \cdots  \int
\Pi_{i=1}^5 
da_i a_i^2  {_b} < isospin| P_{45}^{isospin}| isospin>_a\cr
  & & _b < spin| P_{45}^{spin}| spin>_a < b_1 b_2 b_3 b_4 b_5|
P_{45}^{mom} 
| a_1 a_2 a_3 a_4 a_5 >\cr
  & &  < a|\psi_{12,123}^{1234}>
\end{eqnarray}
The permutation is separated into the 3 spaces: isospin, spin,
and momentum, and  the $ \delta's\cdots$ are the strings of
Kronecker symbols.

The isospin- and spin-matrix elements can be calculated in a standard manner:
\begin{eqnarray}
& &   {_b} < isospin| P_{45}^{isospin}| isospin>_a
 =  \delta_{T T'} \delta_{M_T m_T'} \delta_{ t_{12} t_{12}'}
( - ) ^{ t_{12} + t_{1-4} + 1} \cr
&& \sqrt{ \hat t_3' \hat t_{34} \hat
    t_{1-4} 
\hat t_{56} \hat t_4' \hat t_{56}'}
  \left\{{\begin{array}{*{20}c}
   t_{12} & \frac{1}{2} & t_3'  \\
   \frac{1}{2}  & t_{1-4}  & t_{34}  \\
\end{array}}\right\}
 \left\{{\begin{array}{*{20}c}
   t_3' & {\frac{1}{2}} & t_{1-4}  \\
 {\frac{1}{2}}    & {\frac{1}{2}} & t_{56}  \\
  t_4' & t_{56}' & T  \\
\end{array}}\right\}
\end{eqnarray}
\begin{eqnarray}
& &   {_b} < spin| P_{45}^{spin}| spin>_a
 =   \delta_{ s_{12} s_{12}'}
( - ) ^{ s_{12} + S + 1} \sqrt{ \hat I_3' \hat s_{34} \hat S \hat
    s_{56} 
\hat I_4' \hat s_{56}'}\cr
 & & \left\{{\begin{array}{*{20}c}
   s_{12} & \frac{1}{2} & I_3'  \\
   \frac{1}{2}  & S  & s_{34}  \\
\end{array}}\right\}
 \left\{{\begin{array}{*{20}c}
   I_3' & {\frac{1}{2}} & S  \\
 {\frac{1}{2}}    & {\frac{1}{2}} & s_{56}  \\
  I_4' & s_{56}' & J  \\
\end{array}}\right\}
\end{eqnarray}

For the momentum space part we insert a complete basis and obtain
\begin{eqnarray}
  < b_1 b_2 b_3 b_4 b_5| P_{45}^{mom} & = &  \int d^3 b_1' 
\cdots d^3 b_5'< b_1 b_2 b_3 b_4 b_5| \vec b_1' \vec b_2'\vec b_3'
\vec b_4'\vec b_5'>
   < \vec b_1' \vec b_2'\vec b_3'\vec b_4'\vec b_5'| P_{45}\cr
   &  = &  (\frac{1}{\sqrt{4 \pi}})^5 \int d \hat b_1' \cdots d 
\hat  b_5'
   <  b_1 \hat b_1'   b_2 \hat  b_2' b_3 \hat  b_3' b_4 \hat  b_4' b_5 
\hat  b_5'| P_{45}^{mom}
   \end{eqnarray}

The transposition $ P_{45}^{mom}  $ acting to the left leads to a
state with the same quantum  numbers but different meaning. 
Particles 4 and 5 are interchanged. That state can be reexpressed
again in terms of the old b-state using the relation  between the  Jacobi
   momenta $ \vec b_1,\cdots \vec b_5 $ and the Jacobi momenta with 
particles 4 and 5 interchanged. It results
\begin{eqnarray}
    < \vec b_1 \vec b_2 \vec b_3 \vec b_4 \vec b_5| P_{45}
    &=&   < \vec b_1, \frac{1}{2} ( \vec b_2 - \frac{1}{2} \vec b_3 
-   \vec b_4 - \frac{3}{4} \vec b_5),
     \frac{1}{2} ( -\vec b_2 + \frac{3}{2} \vec b_3 -  \vec b_4   
- \frac{3}{4} \vec b_5),\cr
   & & \frac{1}{2} ( -\vec b_2 - \frac{1}{2} \vec b_3 +  \vec b_4   
- \frac{3}{4} \vec b_5),
    -   ( \vec b_2 + \frac{1}{2} \vec b_3 +  \vec b_4  -
\frac{1}{4} 
\vec b_5)|
   \end{eqnarray}

Consequently
\begin{eqnarray}
& & < b_1 b_2 b_3 b_4 b_5|  P_{45}^{mom} =  (\frac{1}{\sqrt{4 \pi}})^5 
\int d \hat b_1' \cdots d \hat b_5'\cr
& &   < b_1 \hat b_1', \frac{1}{2} ( b_2 \hat b_2' - \frac{1}{2} b_3 
\hat b_3' -  b_4 \hat b_4' - \frac{3}{4}   b_5 \hat b_5'),
\frac{1}{2} ( - b_2 \hat b_2' + \frac{3}{2}  b_3 \hat b_3' - b_4 \hat
b_4' 
- \frac{3}{4}  b_5 \hat b_5'),\cr
  & &   \frac{1}{2} ( - b_2 \hat b_2' - \frac{1}{2}  b_3 \hat b_3' +
b_4 
\hat b_4'   - \frac{3}{4}  b_5 \hat b_5'),
-   (  b_2 \hat b_2' + \frac{1}{2} b_3 \hat b_3' +  b_4 \hat b_4'  
-  \frac{1}{4} b_5 \hat b_5')|\label{87}
\end{eqnarray}

The state $ | \vec b_1 \cdots \vec b_5>$  can be reexpressed in terms
of the state $ | \vec a_1 \cdots \vec a_5>$  as
\begin{eqnarray}
  | \vec b_1 \vec b_2 \vec b_3 \vec b_4  \vec b_5> = |  \vec b_1 ,
  \frac{2}{3} 
( \vec b_2 - \vec b_3), - \frac{1}{2} ( 2 \vec b_2 + \vec b_3),
   \vec b_4, \vec b_5>\label{91}
   \end{eqnarray}

Consequently using (\ref{87}) and ( \ref{91}) one obtains
\begin{eqnarray}
& & < b_1 b_2 b_3 b_4 b_5|  P_{45}^{mom} \int \Pi_{i=1}^5 da_i a_i^2 
| a_1 a_2 a_3 a_4 a_5 > <  a_1 a_2 a_3 a_4 a_5 | \psi_{12,123}^{1234}>\cr
  & = &   \int d \hat  b_1' \cdots d \hat b_5'  <  b_1,  \frac{1}{3} 
| 2 b_2 \hat b_2' - 2  b_3 \hat b_3'|,
   \frac{1}{2} |  \frac{1}{2} b_2 \hat b_2' + \frac{1}{4}  b_3 \hat
   b_3' 
-  \frac{3}{4}  b_4  \hat b_4' - \frac{9}{8}   b_5 \hat b_5')|,\cr
  & &   \frac{1}{2} | - \frac{1}{2} b_2  \hat b_2' - \frac{1}{2}  b_3 
\hat b_3' +  b_4 \hat b_4' - \frac{3}{4} b_5 \hat b_5')|,\cr
  & &    | b-2 \hat b_2' + \frac{1}{2}  b_3 \hat b_3' +  b_4 \hat b_4'   
- \frac{1}{4} b_5 \hat b_5' | | \psi_{12,123}^{1234}>( \frac{1}{\sqrt{4 \pi}})^5
\end{eqnarray}

The angular integration over $ \hat b_1'$ yields directly $ 4 \pi $. 
Further one can put $ \hat b_2' $ into the z-direction
  and $\hat b_3'$  into the x-z plane. Therefore
\begin{eqnarray}
& & < b_1 b_2 b_3 b_4 b_5|  P_{45}^{mom} \int \Pi_{i=1}^5 da_i a_i^2 
| a_1 a_2 a_3 a_4 a_5 > <  a_1 a_2 a_3 a_4 a_5 | \psi_{12,123}^{1234}>\cr
 & = &  \sqrt{\pi}  \int  d cos \theta_3' d \hat  b_4' d \hat b_5'
    <  b_1,  \frac{1}{3} | 2 b_2 \hat b_2' - 2  b_3 \hat b_3'|,
     \frac{1}{2} |  \frac{1}{2} b_2 \hat b_2' + \frac{1}{4} b_3
\hat b_3' 
-  \frac{3}{4}   b_4 \hat b_4' - \frac{9}{8}   b_5 \hat b_5')|,\cr
  & &   \frac{1}{2} | - \frac{1}{2} b_2  \hat b_2' - \frac{1}{2} b_3
\hat b_3' 
+  b_4 \hat b_4' - \frac{3}{4} b_5 \hat b_5')|,
      | b_2 \hat b_2' + \frac{1}{2} b_3 \hat b_3' +  b_4 \hat b_4'   
- \frac{1}{4} b_5 \hat b_5'| | \psi_{12,123}^{1234}>
\end{eqnarray}

This is an example where a 4-dimensional interpolation appears necessary.
Therefore if the momentum magnitudes are discretized choosing for each
one a certain grid, interpolations are inevitable.

In the 3- and 4-nucleon problems cubic Hermitean spline interpolation 
turned out to be very efficient \cite{hueber}. 
Thus a 5-dimensional interpolation, for instance,  has the form
\begin{eqnarray}
&&f_{ijklm} ( a_1 a_2 a_3 a_4 a_5)  =  \cr
&& \sum_{r=0}^3 \sum_{s = 0}^3 \sum_{t=0}^3 \sum_{u=0}^3 
\sum_{v=0}^3 S_r( a_1 )S_s( a_2 )S_t( a_3 )S_u( a_4 )S_v( a_5 )
     f( a_{1r} a_{2s}a_{3t}a_{4u}a_{5v})
\label{80}
\end{eqnarray}
where $ ijklm $ denotes the 5-dimensional cubus around the point 
$ a_1 a_2 a_3 a_4 a_5 $ and $ a_{ik} $ denotes the grid points for 
the variable $a_i$.

For $ a_1 $,  for instance,
   the 4 grid points related to $a_i$ are   $ a_{10} < a_{11} \leq a_1 
\leq  a_{12} < a_{13} $
 and similar for the other variables. Here for the sake of a simpler 
notation we renumbered the grid points in that context.

Further $ f $ is the function to be interpolated
   and $ f_{ijklm} $ the interpolating one.  The spline functions 
are given in Appendix \ref{appendix_b} and the conditions underlying that form
(\ref{80}) can be found in \cite{hueber}.

The coupled set (\ref{70})-(\ref{73}) in a matrix notation has  the 
schematic structure
\begin{eqnarray}
        \eta ( E) \psi = K(E) \psi
\end{eqnarray}
where $E$ is the searched for energy eigenvalue at which the auxiliary 
kernel eigenvalue $\eta(E)=1$. For 3- and 4-nucleon bound states
   a Lanczos type algorithm turned out to be very 
efficient  \cite{stadler,nogga1}. Starting from an arbitrary 
initial $\psi = \psi_0 $
 one generates by consecutive applications of $K$ a sequence of
 amplitudes $\psi_n$, which after orthogonalisation form a basis into which
    $\psi$ is expanded. It turn out that a reasonably small number 
of K-applications (of the order of 10-20) is sufficient, which leads
to an algebraic eigenvalue problem
 of rather low dimension. Then the  energy is varied such that one 
reaches $ \eta(E)=1 $.

If one regards the sub clustering underlying the 5 amplitudes only 
two of them, $ \psi^{1234}_{12,123} $ and $ \psi^{1234}_{12;12,34} $,
  are related to the very approximative effective 3-body model of an 
inert $ \alpha $-core and two neutrons. The total isospin
 quantum numbers of the $^6He $ ground state are $T=1$ and $M_T =-1$ 
(if we define the magnetic isospin quantum number of the neutron 
as $-\frac{1}{2}$).

Now even these two amplitudes depending on the Jacobi momenta of type 
a and b,  (\ref{74}) and (\ref{106}), respectively, are not restricted to
  $ t_4 =0 $ and $ t_{56}=1$ in case of $ \psi^{1234}_{12,123}$ but 
also allow $ t_4 =1 $ and $ t_{56}=1 $, $ t_4 =2 $ and $ t_{56}=1 $
 and $ t_4 =1 $ and $t_{56}=0 $. Similarily the "deuteron-deuteron"
 like substructures of $  \psi^{1234}_{12;12,34}$ are  not restricted to
  $ t_{12}=t_{34}=0$ and $ t_{56}=1$ but also other two-body isospins
 are allowed which do not built up a $ t=0 $ $\alpha$-core. The amplitudes
    $ \psi^{123,45}_{12,123}$ going with Jacobi momenta of type c,
 (\ref{6.4}), refer to a 3-body together with a two-body sub
 clustering, which is
 also  not present in the effective 3-body model. The linear
 combinations  $ \psi^{125,34}_{12;12,34} + \psi^{345,12}_{12;12,34}$ refer
  again to a 3-body together  with a 2-body sub clustering, where 
the underlying fragmentation related to 2-body fragments  differs from
  $ \psi^{123,45}_{12,123}$. And  again this is beyond the effective 
3-body model. Finally  $ \psi^{12,34,56}_{12;12,34}$ allows for several
 additional 2-body sub clusters which are not contained in the 
effective 3-body model, either.

If one would add another step in the Yakubovsky scheme, namely  to 
2-body fragmentations $ a_2 $,  the resulting  amplitudes  
$ \psi^{a_2}_{a_5,a_4,a_3}$
 would point to 5-body subclusters together with a single  nucleon or 
to two 3-body subclusters (like $^3H-^3H$) in addition to  
4-body and  2-body subclusters. Now all those additional structures
are of course also generated by the coupled system
(\ref{70})-(\ref{73})  ending with $a_3$ fragmentation, which we presented.
 In other words that Yakubovsky scheme is complete and delivers an 
exact description of the 6-nucleon problem.

\section{Summary}
\label{sec_five}

The Yakubovsky equations have been derived long time ago by 
O. A. Yakubovsky \cite{yakubovsky}. Therefore the application
presented here could have been given also long time ago. It is,
however, only now after the experiences with 3- and 4-nucleon 
problems in the Faddeev-Yakubovsky schemes that the technical
 expertise has been developed in the last decades and the very strong 
increase  of computer  power just recently achieved allows to attack
the 6-body problem in that exact formulation. Therefore we felt it is 
timely to work out that scheme for that system. Another argument is
the development of nuclear forces in a systematic manner in the realm 
of effective field theory and based on chiral symmetry. Two-, three-, 
and four-nucleon forces have been derived
  consistently to each other and they are waiting to be applied in 
light nuclear systems and checked against nature. Several tests in 
that spirit already appeared \cite{ncsm,gfmc,nogga1,gitter} but for 
the purpose of benchmarking the exact approach in the Yakubovsky 
scheme is strongly recommended.

Here we restricted the formulation to two-nucleon forces only but 
the inclusion of three-nucleon forces can easily be done like 
pioneered in \cite{kamada2}.

In section \ref{sec_two} we used the general 
basic and standard formulation, 
where $a_n$ points to  n-body fragmentations for a system of $ N > n$ particles.

We worked out that scheme ending with $a_3$ in the spirit of the 
usually applied though approximate effective $ \alpha -n -n $ 3-body model.

Though the step to $ a_2$ could be easily done, ending with $ a_3$
still includes exactly the whole dynamics.

The Pauli principle is then exactly incorporated in section
\ref{sec_three} leading to a set of 5 coupled equations for 5 
 independent Yakubovsky components, which built up the total state.

The technical performance in a partial wave decomposition is only 
touched in section \ref{sec_four}. The 5 different Jacobi momenta 
as well as a necessary 
 multi-dimensional interpolation scheme, like modified cubic
 Hermitean splines, are given.  For solving the high dimensional
 energy eigenvalue problem of the 5 coupled equations  we point 
to  the Lanczos type algorithm, which turned out to be very efficient 
in the 3- and 4-nucleon problem.  It remains to work out the partial 
wave projected kernels, which is straightforward and can be carried 
through along the lines cited above. An example for that is presented.

We expect that on the most modern supercomputers with parallel 
architecture this formulation can be numerically mastered.

\begin{acknowledgments}
This work was partially supported 
 by the Polish 2008-2011 science 
funds as the research project No. N N202 077435 
and by the Helmholtz
Association through funds provided to the virtual institute ``Spin
and strong QCD''(VH-VI-231). 
\end{acknowledgments}

\appendix
\section{Jacobi momenta related to the independent 
Yakubovsky components}
\label{appendix_a}

Here we display various Jacobi momenta related to the independent 
Yakubovsky components.

To $ \psi_{12; 12,34}^{ 1234} $ belongs
\begin{eqnarray}
    \vec b_1 & = &  \frac{1}{2} ( \vec k_1 - \vec k_2) = \vec a_1\cr
 \vec b_2  & = &  \frac{1}{2} (  \vec k_3 - \vec k_4 )\cr
   \vec b_3  & = &  \frac{1}{2} ( \vec k_1 +  \vec k_2 - \vec k_3 
- \vec k_4 )\cr
  \vec b_4  & = &  \frac{1}{2} ( \vec k_5 - \vec k_6) = \vec a_4\cr
    \vec b_5  & = &  \frac{1}{3} ( 2 ( \vec k_5 + \vec k_6)  - \vec
    k_1 
- \vec k_2 -\vec k_3 - \vec k_4 )= \vec a_5\label{106}
\end{eqnarray}

The individual momenta are expressed in terms of those Jacobi momenta:
\begin{eqnarray}
  \vec k_1 & = &  \vec b_1 +   \frac{1}{2}\vec b_3 - \frac{1}{4}\vec b_5\cr
  \vec k_2  & = &  - \vec b_1 +   \frac{1}{2}\vec b_3 - \frac{1}{4}\vec b_5\cr
  \vec k_3  & = &  \vec b_2 -  \frac{1}{2}\vec b_3 - \frac{1}{4}\vec b_5\cr
  \vec k_4  & = &  - \vec b_2 - \frac{1}{2}\vec b_3 - \frac{1}{4}\vec b_5\cr
  \vec k_5  & = &  \vec b_4 + \frac{1}{2}\vec b_5
  \end{eqnarray}

The kinetic energy is
\begin{eqnarray}
   \sum_{i=1}^{6}\frac{k_i^2}{2m} = \frac{1}{2 m} ( 2 b_1^2 + 2 b_2^2 
+  b_3^2 + 2 b_4^2 + \frac{3}{4} b_5^2)
\end{eqnarray}

To $ \psi_{12,123}^{ 123,45} $ belongs
\begin{eqnarray}
 \vec c_1 & = &  \frac{1}{2} ( \vec k_1 - \vec k_2)= \vec a_1\cr
 \vec c_2  & = &  \frac{1}{3} ( 2 \vec k_3 - \vec k_1- \vec k_2 ) = \vec a_2\cr
 \vec c_3  & = &  \frac{1}{2} ( \vec k_4  - \vec k_5  )\cr
 \vec c_4  & = &  \frac{1}{3} ( \vec k_4 + \vec k_5 - 2\vec k_6)\cr
 \vec c_5  & = &  \frac{1}{2} (   \vec k_4 + \vec k_5 + \vec k_6  
- \vec k_1 - \vec k_2 -\vec k_3)\label{6.4}
\end{eqnarray}
or
\begin{eqnarray}
   \vec k_1 & = &  \vec c_1 -  \frac{1}{2}\vec c_2 - \frac{1}{3}\vec c_5\cr
  \vec k_2  & = &  - \vec c_1 -   \frac{1}{2}\vec c_2 - \frac{1}{3}\vec c_5\cr
  \vec k_3  & = &  \vec c_2 -  \frac{1}{3}\vec c_5\cr
  \vec k_4  & = &   \vec c_3 + \frac{1}{2}\vec c_4 +\frac{1}{3}\vec c_5 \cr
  \vec k_5  & = &  - \vec c_3 + \frac{1}{2}\vec c_4 +\frac{1}{3}\vec c_5 \cr
 \vec k_6  & = &  - \vec c_4  +\frac{1}{3}\vec c_5
\end{eqnarray}
and the kinetic energy is
\begin{eqnarray}
  \sum_{i=1}^{6}\frac{k_i^2}{2m} =  \frac{ 1}{m} c_1^2 + \frac{3}{4m}
  c_2^2 
+  \frac{ 1}{m} c_3^2 + \frac{3}{4m} c_4^2 + \frac{1}{3m} c_5^2
\end{eqnarray}

To $ \psi_{12; 12,34}^{ 125,34} $ belongs
\begin{eqnarray}
    \vec d_1 & = &  \frac{1}{2} ( \vec k_1 - \vec k_2) = \vec a_1\cr
 \vec d_2  & = &  \frac{1}{3} ( 2 \vec k_5 - \vec k_1- \vec k_2 )\cr
   \vec d_3  & = &  \frac{1}{2} ( \vec k_3  - \vec k_4  ) = \vec b_2\cr
  \vec d_4  & = &  \frac{1}{3} ( \vec k_3 + \vec k_4 - 2\vec k_6)\cr
    \vec d_5  & = &  \frac{1}{3} (   \vec k_3 + \vec k_4 + \vec k_6  
- \vec k_1 - \vec k_2 -\vec k_5)\label{112}
\end{eqnarray}
or
\begin{eqnarray}
   \vec k_1 & = &  \vec d_1 -  \frac{1}{2}\vec d_2 - \frac{1}{3}\vec d_5\cr
  \vec k_2  & = &  - \vec d_1 -   \frac{1}{2}\vec d_2 - \frac{1}{3}\vec d_5\cr
  \vec k_3  & = &  \vec d_3 +  \frac{1}{2}\vec d_4 + \frac{1}{3}\vec d_5\cr
  \vec k_4  & = &   - \vec d_3 + \frac{1}{2}\vec d_4 +\frac{1}{3}\vec d_5 \cr
  \vec k_5  & = &   \vec d_2 - \frac{1}{3}\vec d_5 \cr
 \vec k_6  & = &  - \vec d_4  +\frac{1}{3}\vec d_5
\end{eqnarray}
and the kinetic energy is
\begin{eqnarray}
  \sum_{i=1}^{6}\frac{k_i^2}{2m} =  \frac{ 1}{m} d_1^2 + \frac{3}{4m}
  d_2^2 
+  \frac{ 1}{m} d_3^2 + \frac{3}{4m} d_4^2 + \frac{1}{3m} d_5^2
\end{eqnarray}

To $ \psi_{12; 12,34}^{ 345,12} $ belongs
\begin{eqnarray}
    \vec e_1 & = &  \frac{1}{2} ( \vec k_3 - \vec k_4) = \vec b_2\cr
 \vec e_2  & = &  \frac{1}{3} ( 2 \vec k_5 - \vec k_3- \vec k_4 )\cr
   \vec e_3  & = &  \frac{1}{2} ( \vec k_1  - \vec k_2  ) = \vec a_1\cr
  \vec e_4  & = &  \frac{1}{3} ( \vec k_1 + \vec k_2 - 2\vec k_6)\cr
    \vec e_5  & = &  \frac{1}{3} (   \vec k_1 + \vec k_2 + \vec k_6  
- \vec k_3 - \vec k_4 -\vec k_5)\label{115}
\end{eqnarray}
or
\begin{eqnarray}
   \vec k_1 & = &  \vec e_3 +  \frac{1}{2}\vec e_4 + \frac{1}{3}\vec e_5\cr
  \vec k_2  & = &  - \vec e_3 +   \frac{1}{2}\vec e_4 + \frac{1}{3}\vec e_5\cr
  \vec k_3  & = &  \vec e_1 -  \frac{1}{2}\vec e_2 - \frac{1}{3}\vec e_5\cr
  \vec k_4  & = &   - \vec e_1 - \frac{1}{2}\vec e_2 - \frac{1}{3}\vec e_5 \cr
  \vec k_5  & = &   \vec e_2 - \frac{1}{3}\vec e_5 \cr
 \vec k_6  & = &  - \vec e_4  + \frac{1}{3}\vec e_5
\end{eqnarray}
and the kinetic energy is
\begin{eqnarray}
  \sum_{i=1}^{6}\frac{k_i^2}{2m} =  \frac{ 1}{m} e_1^2 + \frac{3}{4m} e_2^2 +  \frac{ 1}{m} e_3^2 + \frac{3}{4m} e_4^2 + \frac{1}{3m} e_5^2
\end{eqnarray}

To $ \psi_{12; 12,34}^{12,34,56} $ belongs
\begin{eqnarray}
    \vec f_1 & = &  \frac{1}{2} ( \vec k_1 - \vec k_2) = \vec a_1\cr
 \vec f_2  & = &  \frac{1}{2} (  \vec k_3 - \vec k_4 ) = \vec b_2\cr
   \vec f_3  & = &  \frac{1}{2} ( \vec k_5  - \vec k_6  ) = \vec a_4\cr
  \vec f_4  & = &  \frac{1}{2} ( \vec k_1 + \vec k_2 - \vec k_3 - \vec k_4)\cr
    \vec f_5  & = &  \frac{1}{3} ( 2(   \vec k_5 +  \vec k_6)   
- \vec k_1 - \vec k_2 -\vec k_3 - \vec k_4) = b_5\label{118}
\end{eqnarray}
or
\begin{eqnarray}
   \vec k_1 & = & \vec f_1 +  \frac{1}{2}\vec f_4  -  \frac{1}{4}\vec f_5\cr
  \vec k_2  & = & - \vec f_1 +  \frac{1}{2}\vec f_4  -  \frac{1}{4}\vec f_5\cr
  \vec k_3  & = &  \vec f_2 -  \frac{1}{2}\vec f_4 - \frac{1}{4}\vec f_5\cr
  \vec k_4  & = &   - \vec f_2 - \frac{1}{2}\vec f_4 - \frac{1}{4}\vec f_5 \cr
  \vec k_5  & = &   \vec f_3 + \frac{1}{2}\vec f_5 \cr
 \vec k_6  & = &  - \vec f_3  + \frac{1}{2}\vec f_5
\end{eqnarray}
and the kinetic energy is
\begin{eqnarray}
  \sum_{i=1}^{6}\frac{k_i^2}{2m} =  \frac{ 1}{m} f_1^2 + \frac{1}{m}
  f_2^2 
+  \frac{ 1}{m} f_3^2 + \frac{1}{4m} f_4^2 + \frac{3}{8m} f_5^2
\end{eqnarray}

\section{Modified spline functions}
\label{appendix_b}

Choosing four grid points $x_0$, $x_1$, $x_2$ , and $x_3$  such 
that $ x_1 \leq x \leq x_2 $ the modified spline functions \cite{hueber} are
\begin{eqnarray}
S_0 (x) & = &   - \phi_3 (x) \frac{ x_2 - x_1}{ x_1 - x_0} \frac{1}
{ x_2 - x_0}\cr
S_1 ( x) & = &  \phi_1(x) + \phi_3 (x) (\frac{ x_2 - x_1}{ x_1 - x_0} 
- \frac{ x_1 - x_0}{ x_2 - x_1}) \frac{1}{ x_2 - x_0} -
\phi_4 (x) \frac{ x_3 - x_2}{ x_2 - x_1} \frac{1}{ x_3 - x_1}\cr
S_2(x) & = &   \phi_2(x) + \phi_3 (x) \frac{ x_1 - x_0}{ x_2 - x_1} 
\frac{1}{ x_2 - x_0} + \phi_4 (x) (\frac{ x_3 - x_2}{ x_2 - x_1} - 
\frac{ x_2 - x_1}{ x_3 - x_2}) \frac{1}{ x_3 - x_1}\cr
S_3(x) & = &  \phi_4 (x) \frac{ x_2 - x_1}{ x_3 - x_2} \frac{1}{ x_3 - x_1}
\end{eqnarray}
with
\begin{eqnarray}
\phi_1(x) & = &  \frac{ ( x_2 -x)^2}{ x_2 - x_1)^3} ( ( x_2 - x_1) 
+ 2 ( x - x_1))\cr
\phi_2(x) & = &  \frac{ ( x_1 -x)^2}{ x_2 - x_1)^3} ( ( x_2 - x_1) 
+ 2 ( x_2 - x))\cr
\phi_3(x)  & = &  \frac{ ( x-x_1) ( x-x_2)^2}{ (x_2 - x_1)^2}\cr
 \phi_4(x)  & = &  \frac{ ( x-x_1)^2 ( x-x_2)}{ (x_2 - x_1)^2}
\end{eqnarray}


\begin{thebibliography}{99}



\bibitem{rep} M. V. Zhukov et al., Phys. Rep. {\bf 231} (1993) 151.

\bibitem{epelbaum} E. Epelbaum, 
Prog. Part. Nucl. Phys. {\bf 57}, 654 (2006), nucl-th/0509032.

\bibitem{ncsm}A. Nogga, arXiv:nucl-th/0611081; 
 A. Nogga, P. Navratil, B.R. Barrett, J.P. Vary, Phys. Rev. C {\bf 73}, 064002
(2006); 
 T.C. Luu, P. Navratil, A. Nogga, arXiv:nucl-th/0412109.

\bibitem{gfmc} B. S. Pudliner, V. R. Pandharipande, J. Carlson, 
and R. B. Wiringa, Phys. Rev. Lett. {\bf 74}, 4396  (1995).

\bibitem{varga} K. Varga, Y. Suzuki, and Y. Ohbaysi, Phys. Rev C {\bf 50},
  189 (1994).

\bibitem{evgeny}  E. Epelbaum, private communication.

\bibitem{kamada1}  H. Kamada, W. Gloeckle, Nucl.  Phys. {\bf A 548}, 205 (1992).

\bibitem{kamada2} W. Gloeckle,  H. Kamada, Nucl. Phys. {\bf A 560}, 541 (1993).

\bibitem{kamada3}  H. Kamada, W. Gloeckle,  Phys. Lett {\bf B 292}, 1 (1992).

\bibitem{kamada4} W. Gloeckle,  H. Kamada, Phys. Rev. Lett. {\bf 71}, 
971 (1993).

\bibitem{nogga}   A. Nogga,  H. Kamada, W. Gloeckle, Phys. Rev
  Lett. {\bf 85}, 944 (2000).

\bibitem{nogga1}   A. Nogga,  H. Kamada, W. Gloeckle, 
B. R. Barrett,  Phys. Rev C {\bf 65}, 054003 (2002).

\bibitem{nogga2}   D. Rozp\c edzik et al., Acta Phys. Polonica {\bf B 37}, 2889
  (2006).

\bibitem{yakubovsky}   O. A. Yakubovsky, Sov. J. Nucl. Phys. {\bf 5}, 
937 (1967).

\bibitem{mybook} W. Gloeckle,  The Quantum Mechanical Few-Body
  Problem, Springer Verlag, 1983.

\bibitem{witala}   H. Wita{\l}a, W. Gloeckle, Eur. Phys. J. {\bf A
  37}, 87 (2008).

\bibitem{hueber}   D. Hueber et al., Few-Body Systems  {\bf 22}, 107 (1997).

\bibitem{stadler}   A. Stadler, W. Gloeckle, P. U. Sauer,   
Phys. Rev C {\bf 44}, 2319 (1991).

\bibitem{gitter}   E. Epelbaum, H. Krebs, D. Lee, Ulf-G. Meissner, 
Eur. Phys. J. {\bf A 45}, 335 (2010); Phys. Rev. Lett. {\bf 104}, 142501 (2010).

\end{thebibliography}
\end{document}